\documentclass[conference]{IEEEtran}
\usepackage{balance}
\usepackage{amsmath,amssymb,amsfonts}
\usepackage{algorithmic}
\usepackage{graphicx}
\usepackage{textcomp}
\def\BibTeX{{\rm B\kern-.05em{\sc i\kern-.025em b}\kern-.08em
		T\kern-.1667em\lower.7ex\hbox{E}\kern-.125emX}}
\usepackage{url}
\usepackage{booktabs}
\usepackage{float}
\usepackage{placeins}
\restylefloat{table}
\usepackage{supertabular,booktabs}
\usepackage[nocompress]{cite}
\usepackage{makecell}
\usepackage{url}

\usepackage[dvipsnames]{xcolor}

\begin{document}
	\title{Exploitation of Human~Trust, Curiosity and Ignorance by Malware}
	\author{	\IEEEauthorblockN{Sundar~Krishnan}
				\IEEEauthorblockA{\textit{Department of Computer Science} \\
				\textit{Sam Houston State University}\\
				Huntsville, TX, USA \\
				E-mail: \textit{skrishnan@shsu.edu}} }
	\maketitle

	\begin{abstract}
	Despite defensive advances in the Internet realm, Malware (malicious software) remains a Cybersecurity threat. These days, Malware can be purchased and licensed on the Internet to further customize and deploy. With hundreds of Malware variants discovered every day, organizations and users experience enormous financial losses as cybercriminals steal financial and user data. In this article surveys the human characteristics that are key to the defense chain against Malware. The article starts with the attack models/vectors that humans often fall prey to and their fallouts. Next, analysis of their root cause and suggest preventive measures that may be employed is detailed. The article concludes that while Internet user education, training, awareness can reduce the chances of Malware attacks, it cannot entirely eliminate them.
	\end{abstract}
	

	\section{Introduction}
	Many Internet users are not knowledgeable enough to differentiate between real and malicious online entities. The Internet user pool is rapidly growing across the world. Many other under-developed countries are moving in the developing world, and this shift is bound to bring in more new Internet users. User ignorance not only results in their exploitation but also dramatically impacts other users within social networks or the organization. The curiosity character of a user leads them to explore the Internet without care, thereby exposing them to attack vectors. Although a large proportion of Internet users are aware of the possible Malware exploits, they seem to forget their warnings and lower their guard by trusting websites and URLs (hyperlinks). McAfee Labs researchers documented a record number of Malware samples in 2019 targeting Mobiles, Mac systems, Linux, IoT, Bitcoins, JavaScript, Internet, and Macros \cite{McAfee}. User awareness regarding Malware security concerns is critical but can only spread gradually. Therefore social networks, Internet Service providers, online search engines should be proactive and develop more sophisticated and stringent mechanisms to thwart Malware infections. Safe and secure transmission of online information and robust user’s privacy should be the paramount concern of all players on the Internet. The objective of this paper is to outline the vulnerable role that humans play in the fight against Malware and identify areas for improvement. The author uses multiple approaches to document reasons for human victimization by Malware attacks.
		
	\section{Problem Statement}
	Human Trust, Curiosity and Ignorance are the most unreliable link of the entire Malware defense chain.
	This paper’s objectives are as follows:
	
	\begin{enumerate}
		\item To outline exploitation mechanisms used by Malware authors to trick human ignorance into being a victim of Malware attacks.
		\item To highlight the preventive measures that can be taken to safeguard any Internet user’s privacy space.
	\end{enumerate}
	
	\section{Motivation and Need}
	
	\subsection{Human Trust}
	Lee et at. in 1991 reported that since the development of the Internet, there is considerable evidence that developing supportive interpersonal relationships online are important \cite{Sproull}. They argue that in a networked organization, the focus of attention changes from the relationship between a person and technology to the relationship between a person and other people. Often, Internet users who never know each other can work collectively through information sharing and group communication across the Internet. Many people provide open access to data and information to others on the Internet, implying that certain level of trust may exist because of the information owner’s credit and privacy is at risk.
	
	\begin{figure*}
		\centering
		\includegraphics[width=0.95\linewidth]{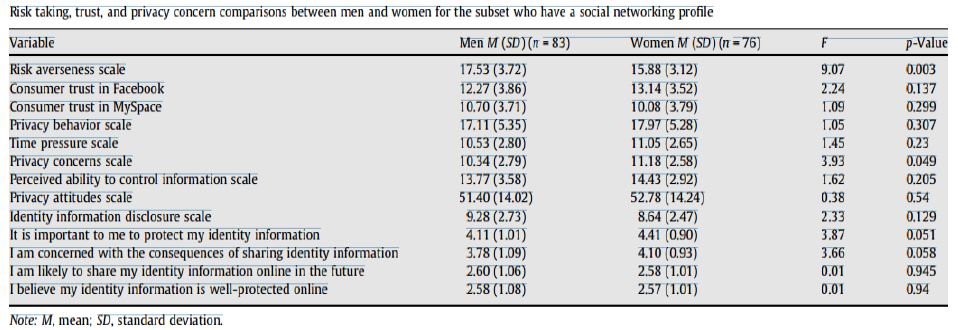}
		\caption{\label{fig:Malware} Risk, Privacy and trust with/without Social Networking profiles \cite{Fogel2009}}
	\end{figure*}
			
	Trust in human-computer interaction (HCI) focuses on users trusting websites, applications, online friends. Users seldom feel that a certain online merchant is untrustworthy. Figure \ref{fig:Malware} shows statistical a significant difference between men and women regarding risk averseness scale (p=0.003) and a borderline significance on privacy concerns scale (p=0.049). Internet users often trust online information, such as reviews and comments during their decision making \cite{Chua_TripAdvisor_reliability}. Few factors that establish trustworthiness of a website tend to be related to showcasing the physical (face-to-face) nature of the organization or its customers such as buildings, any public organization reports, Corporate Social Responsibility (CSR) initiatives, user-reviews, employee testimonials, and few pictures of employees in their work place setting. Social media platforms are becoming increasingly popular news sources as they distribute information from various Internet users and numerous media outlets. An online study \cite{Buchanan2019} of 357 Facebook users found that messages from trusted sources were more likely to be propagated irrespective of propensity risk. This suggests that messages(news) can spread irrespective of it being genuinely true or false just by trusting the message source and recipient. In a study of about 126,000 Twitter stories that were tweeted (discussed/shared) 4.5 million times, the authors found that humans, not robots, were more likely to spread false news as it was more novel than true news. This shows that human fascination for the novelty of news was a catalyst for its propagation irrespective of its genuineness being verifiable. Sterrett et al. \cite{Sterrett2019} studied factors such as the trustworthiness of the person who shares a story and the credibility of the news outlet reporting the story. The authors concluded that a source has a limited effect on the sharing of their news stories while the sharer has a strong impact on attitudes towards a story. Such human behavior coupled with current Internet speeds, its reach, minimal verifications, social pressure to contribute/share and the busy lives of Internet users can explain their implicit trust patterns without cross-verifying messages for their truth. Such behavior can also cause them to indiscriminately share messages with links to malicious URLs and Malware laden attachments thereby creating a perfect petri-dish for Malware authors.
	
	\subsection{Human Ignorance}
	There is a relatively obvious sense in which ignoring information is the right thing to do. Grunwald and Halpen \cite{Grunwald2014} argue that occasionally we are better off ignoring information if any uncertainty of the information is represented by a set of probability measures. If information can be untrue, it's better to ignore and not absorb it rather than blindly following it. They argue, using both a Bayesian and a non-Bayesian analysis that in some situations, you are better off ignoring information if your uncertainty is represented by a set of probability measures. With the developing country’s population getting increasingly hitched on to the Internet, any false information posted on the Internet can easily spread, causing havoc. Also, the literacy level of the user does not determine if they can distinguish a piece of false information from a true one since they start to trust the Internet more and more. For Instance, not all information on Wikipedia \cite{Wiki_disclaimer} is accurate. Readers should be aware of this and not always trust word for word on online information unless backed by some sort of reputation scores.
	
	In the Internet age, ignorance travels as rapidly as great ideas. This illustrates a drawback of the Internet - the only news delivery system we've ever had that has no editor or moderator. Internet users must always remember that what we read on the Internet may not always be true. User ignorance aids cybercrime as there are too many technical features on websites that the user is ignorant about, causing them to pay little attention to the technical fine print before browsing any further on the site. In a study released by IObit \cite{Survey_30}, the mere option to choose "Keep Me Logged In" when accessing their social media accounts can be manna for Malware. Often users tend to quickly click the accept/close button on messages about privacy terms or use of cookies. They do not realize the pitfalls of skipping such information, and neither do they take a moment to educate themselves about it. 
	
	\subsection{Human Curiosity}
	Curiosity is the nutrition of Social Engineering. Wainer, Dabbish and Kraut \cite{Wainer2012} conducted an experiment to examine how eMail message importance, subject line specificity, workload and personal utility influence attention to our emails. Results suggested that uncertainty about message content at the inbox level increases the likelihood of attention to a message. The influence of uncertainty diminished in the face of enhanced task and personal utility cues and increased demand, suggesting that curiosity operates in an intrinsic way in the eMail context.
	
	Seeburger conducted an exploratory project to investigate if and how PlaceTagz (QR Codes) \cite{Seeburger2012} are used by urban dwellers in a real-world deployment. From the findings, it was observed that QR codes, which did not contain any contextual information, piqued the curiosity of users who wondered about the embedded link’s destination and provoked comments in regards to people, place, and technology. As the poet Dorothy Parker stated: “The cure for boredom is curiosity. There is no cure for curiosity”. It is apt to add that “since the Internet is needed to cure boredom, there is bound to be some level of curiosity involved to keep surfers online."
	
	Vishing, Phishing, Baiting, Dumpster Diving, Hoaxing and other social engineering tactics are all related to user curiosity. The first feature of interest in a computer or tablet for a lay user is Internet functionality. People feel safe to click away if they have antivirus installed. Navigational hyperlinks are well hidden and confusing that images like a download button are found multiple times on a page when we only need to use one button for the needed download. The recent Coronavirus (2019-nCoV) outbreak has been presumed to be used by hackers to spread Malware via emails and attachments. The Malware is disguised as legitimate information about Coronavirus. Scientific news of the outbreak along with various conspiracy theories have lured Internet users towards this topic thus helping Malware authors distribute Phishing eMail attachments that can install the AgentTesla Keylogger \cite{Coronavirus_malware}.
	
	\section{Malware attacks via the Internet}
	Malware is a malicious software designed to infect your computer without your knowledge. Malware is most often used to steal personal information, send junk eMail (Spam), or spread more Malware. Given its popularity and ubiquity, the Internet always attracts the attention of malicious entities and the innocent general public. The Internet continues to grow with free information of all sorts such as videos, news, pornography, social discussions, music, pictures, games etc. Usually to avoid boredom and seek information, a Internet user's human traits of curiosity, ignorance, and trust can lure them into returning to the Internet world especially when content is freely available. The quest to (sometimes aimlessly) click around the website to find some information or download a movie/music/software/game (files) can become unknowing addictive. Few Internet users can also be termed "chronic clickers" as they are careless on whet they click (URLs). Web-based Malware is circulated when innocent victim users visit malicious Websites that do their best to lure visitors into staying on their website and spending time (drive-by-download). Malware is well hidden on these websites and is designed to conduct various social engineering and cybercrimes, such as gaining control of the victim system, stealing private information, launching denial-of-service attacks and spamming.
	
	The traditional Malware distribution model is push-based where attackers actively search for and infect the victim systems. Due to security technologies such as firewalls providing a good defense against push-based propagation techniques, Malware distribution has now evolved to a pull-based model, where victims unknowingly visit malicious Websites. Under the pull-based Malware distribution model, the channel for the attack is initiated and established unknowingly by potential victims, which significantly lowers the defense barrier for attackers to cross over.
	
	\section{Malware distribution methods}
	Hackers and Malware authors are by no means always criminal geniuses and are historically not intentional in breaking the law \cite{Eddy2019}. Distribution and delivery is a vital part of any Malware campaign. The sight of high profits and low risks are often factors that lure them into this field. There is also the sub-tribe of them known as hacktivists, who take on this field to protest or communicate their opinion. Unless  they shop on the dark web for ready-to-use Malware exploits, they often have to wait patiently, be crafty, and conduct extensive reconnaissance of their target subjects and the target environment. They have to continually evolve and try to be distinctive in their attack signatures. Here we take a closer look at the mechanisms used by them to bait gullible Internet and computer users.
	
	\subsection{Websites}
	The most common method used by attackers to distribute exploits is through code on webpages, but malicious exploits may also arrive by eMail. When we visit a website with malicious code while using vulnerable software, the exploit may be loaded. It is important to note that some legitimate websites might unknowingly and unwillingly host malicious code in their advertising (Malvertising). This means that if we visit a site that is hosting these malicious advertisements, an attempt to compromise our computer will be made. Compromised websites may have an executable file thet can be disguised under a file/folder icon, leading a hasty user to click on it. Few web browsers have improved in their security watch when loading websites, and these days flag sites that may have a low reputation. However, Internet users may not always pay attention to such warnings. Websites laden with malicious intent often have many pop-ups designed with adware (advertising Malware) to bait a user into clicking the pop-up window.
	
	\subsection{eMails}
	Communication across the Internet used to primarily happen via eMails in the early 90s and it continues to still be the primary mechanism of formal company/user communication. It is worth noting that with the advent of smartphones the last few years, communication between Internet users has moved into social platforms and encrypted messengers. These days, free eMail platforms coupled with large cloud storage have made it a routine must-have for many individuals across the globe. Despite the growth and prominence of mobile messengers and chat apps, eMail is still an integral part of billions of Internet users \cite{Statista_email_projections}. However, spam, misuse, and Phishing have plagued the system in the last decade. Phishing is when attackers send malicious emails containing malicious web links, malicious attachments, and fraudulent data-entry forms, all designed to trick people into falling for a scam \cite{Phishing_def_Proofpoint}. Phishing and other eMail-based attacks were top concerns in the 2018 Internet Crime Report issued by the U.S. Federal Bureau of Investigation’s Internet Crime Complaint Center (IC3) \cite{FBI_IC3}. The report pegs the financial losses of victims to millions of dollars in 2018 alone. Malware distribution campaign uses messages masquerading as UPS delivery notification emails. It uses the eMail subject “UPS Delivery Notification Tracking Number”. Another type of Malware known as Cryptolocker targets personal and professional computers, and quickly scrambles or encrypts your data. The phony eMail appears to be from FedEx or UPS.
	
	A piece of Malware known as Fort Disco was designed to launch brute-force password guessing attacks against websites built with popular content management systems like WordPress and Joomla was found to also attack eMail and FTP servers. Once it infects a computer, the Malware periodically connects to a command and control (C\&C) server to retrieve instructions, which usually include a list of thousands of websites to target and a password that should be tried to access their administrator accounts \cite{Mimoso2013}. 
	With eMail mostly offered through web browsers, use of security controls such as end-to-end encryption, two-factor authentication, device whitelisting, etc. can be voided if the eMail user is tricked into clicking a link that the Phishing author has designed. All eMail users need to make sure that they never click blindly on the links within emails unless they know who is sending it, the sender's eMail address, any security flags from the eMail client application (such as spam, junk, phishing), and checking the target URL destination by hovering above the URL itself. Most of the eMail websites and client-side applications have measures to disable all hyperlinks on eMail body, but curiosity and trust sometimes get the user to enable them and fall victim to Malware.
	
	\subsection{Botnets}
	Botnets are collections of Malware infected Internet machine hosts (“bots”) that through Malware infection have fallen under the control of a single entity (“botmaster”). Botnets are the most common vehicle of cyber-criminal activity. They are used for spamming, phishing, denial of service attacks, brute-force cracking, stealing private information, and cyber warfare. Botnets commonly scan large segments of the Internet’s address space for various reasons, such as infecting or compromising hosts, recruiting hosts into a botnet, or collating a list of future targets. The website Songsrpeople.com looks a lot like other amateur-video sites with advertisements along with content. But Web-security investigators at White Ops contend that most of the site's visitors aren't people \cite{ChristopherWallStreet2013}. Rather, they are computer-generated visitors, or "bots," designed to fool advertisers into paying for the traffic. White Ops discovered that more than 30\% of the visitors to the education portal Education.com were robots. In the past month, the site received about four million unique views, according to Quantcast \cite{Caroline2013}.
	
	Bots can be programmed to assume a human persona and are perfect for clicking on online social media buttons like "likes" or "votes". In 2009, bots skewed the online voting in Time magazine's Time 100 poll will the No.1 vote getter turning out to be a 21-year-old founder of "4Chan" - a bulletin board for hackers \cite{Bot_4Chan_1}, \cite{Bot_4Chan_2}.	Russian-linked bots were also in the news for their alleged role in influencing social media during the 2016 U.S. presidential election \cite{Bot_4Chan_1}, \cite{Bot_2016_Election_1}, \cite{Bot_2016_Election_2}, \cite{Bot_2016_Election_3}, \cite{Bot_2016_Election_4}. Agarwal et al. \cite{Agarwal_Bot_2017} study the use of botnets and concluded that networked computers running botnets had been extensively used during the 2014 Crimean Water Crisis and the 2015 Dragoon Ride Exercise. They also concluded that the behavior of these botnets have become increasingly sophisticated over time, both from the perspective of information dissemination as well as their coordination. A notorious botnet "Cutwail botnet" known for sending out millions of spam messages that spread banking Malware and other threats has turned to the Magnitude toolkit to help bolster its automated attack campaigns \cite{Westervelt_banking}. Often the victims of Bots are ignorant that their machine is compromised due to the sophistication and elusiveness of the botnets. This also shows that Cybercriminals quickly adjust their operation to maintain attack continuity and go undetected. 
	
	\subsection{Social Media and Social Engineering}
	Facebook walls, twitter tweets, and other social networking sites are the common hunting grounds for Malware attackers. The potential to cascade and infect other users is high due to design of such networking sites. A new variant of the Dorkbot Malware was discovered spreading through Facebook's internal chat service, in an attempt by cybercriminals to harvest users' passwords. Dorkbot is capable of updating itself and can prevent antivirus software from running security updates. It can also download software from a command~-and~-control (C\&C) server for distributing spam and ransomware \cite{CSO_Facebook}.Another case of Malware via video is rapidly spreading via Facebook to Google Chrome users, at the rate of about 40,000 per hour. If users take the bait, the Malware plugin can access all information stored in their browsers, including accounts with saved passwords, and it hijacks users’ Facebook accounts and spreads via the social network \cite{David_2013}.
	
	\subsection{Smartphone Apps}
	As smartphones and tablets powered by mobile operating systems continue to take over the world, Malware makers see vast opportunities. Since few of these mobile devices are based on open platforms, the user is often provided with little security privileges to fully take over the security of these systems. Reports of Malware targeting smartphones continue to grow and Android is a particular focus since its ecosystem is not controlled as tightly as iOS, Blackberry or Windows Phone. This is a big problem as Android happens to also be the most popular mobile platform on the planet. As per a 2013 Juniper Research findings that more than 80\% of the total enterprise and consumer-owned smartphones remain unprotected from Malware attacks \cite{Juniper2013}. Pop-up advertisements that are seemingly benign have found their way into iPhone and Android operating systems and can act as click-baits for lay users. Advertisements such as “congratulations”, "you won", etc. promising prizes like free Amazon gift cards, iPhones or cash in exchange for clicking on the message or submitting personal information often instead inject Malware \cite{Symantec_PopUp_Adware}.
	
	\subsection{Newbie and Script Kiddies as Malware Creators}
	In 2012, Blackhole Malware exploit kit was sold by its creator for an annual license of \$1,500 or could be rented from its creator for \$200 for one week's use, among other price plans. Malware creators do not have to be very technical to operate them since there was a manual to create other custom malwares that came with this Malware \cite{BBC_Blackhole}.
		
	\subsection{Drop Bait}
	Of recent, various storage media such as USB thumb drives have become cheaper, leading to an average criminal to infect or Trojanize the media. The curious finder of such media has a high probability of inserting the device into their computer. Matthew et al. \cite{Matthew_Campus_USB_Drives} conducted a controlled experiment in which 297 infected USB flash drives were dropped on a large university campus. They concluded that this mechanism of attack was effective with an estimated success rate of 48\% and was expeditious as the first drive connected to the campus network in less than six minutes. Thus attackers would have no problem spreading Malware in an organization or public space by simply dropping an infected USB drive.
	
	\subsection{Dark Web}
	Surface Web is the Internet or World Wide Web that we regularly use and is well indexed by various search engines. The Dark Web or "Deep Web" is attributed to the World Wide Web content that isn't indexed by search engines due to overlay networks that require specific software, VPN configurations, or authorization for access. The Dark Web is intentionally hidden, anonymous, and widely known for illicit activities \cite{Pieter_MalwareBytes_DarkWeb}. Specific browsers like Tor are required to access websites on the Dark Web. Most of the darkweb navigation can be made anonymous, taking privacy to great lengths. Unfortunately, websites hosted on the Dark Web often contain anonymous message boards, online sale of Malware, online marketplaces for drugs, exchanges for stolen financial and private data, Child (and other illegal) porn, Bitcoin exchanges, and other illegal/nefarious content. Michael McGuire of the University of Surrey found that the Dark Web is a seller’s market for custom Malware with four in 10 Dark Web vendors selling targeted hacking services aimed at FTSE 100 and Fortune 500 businesses \cite{Michael_Bromium_2019}.

	\section{Malware Exploit Techniques and Payloads}
	In 2011, Lockheed Martin released a paper defining a "Cyber Kill Chain" model that outlines the steps used by cyber attackers in today’s cyber-based attacks against their targets \cite{Lockheed_Kill_chain}. Various versions of the “Cyber Kill Chain” have since been released like the including AT\&T’s Internal Cyber Kill Chain Model and the Unified Kill Chain \cite{SANS_kill_chain}. The step 5 of the "Cyber Kill Chain" model explains the "Installation" phase wherein the attacker installs Malware on the victim's computer/network. Malware needs a way to spread and achieve its intended goal. The payload is a delivery system that carries the malicious code to the victim. Few known Malware exploit techniques and payloads are outlined below;

		\begin{enumerate}
			\item Adding malicious code to hundreds of thousands of legitimate websites, which then copied Malware to visitors computers
			\item Creating hyperlinks in spam messages to specially created sites that infected PCs
			\item Fake antivirus software that falsely claimed the PC was infected and urged the user to pay a fee to remove viruses
			\item Cryptocurrency miners secretively use the victim's compute power for cryptocurrency mining.
			\item Trojans that attempted to steal financial records stored on the PC
			\item The ZeroAccess rootkit, which downloaded other software that hijacked the PC for use in a botnet - a facility used to overwhelm websites with traffic and force them offline.
			\item Spyware like Key loggers are malicious code that record of what was typed on the PC and then.
			\item Ransomware that attempts to blackmail the PC owner into paying a ransom for the release of their computer files.
			\item A vulnerability is effectively an error in the code or the logic of operation that has gone unnoticed by the developer or vendor. Vulnerabilities both that of application and Operating System are often the go-to mechanisms by Malware authors.
			\item Exploit kits are malicious toolkits with prewritten code that attackers use to search for software vulnerabilities on a target’s computer or mobile device. Such toolkits can also be found for sale on the darkweb.
		\end{enumerate}
	
	Attackers usually use a combination of exploits against different software to gain access to the computer. An exploit detection may be triggered by the user's antivirus software when the user visits a website that contains malicious exploit code - even if the user was not using the vulnerable software being targeted. This does not mean that the user had been compromised. It means that an attempt to compromise the user's computer has been made \cite{Microsoft_Exploits}.
	
	\section{Proposed Solutions}
	The following are few solutions to combat Ignorance, Trust, and Curiosity among Internet users. They may be used in tandem with each other and are by no means a whole perfect solution themselves.\\
	
	1. Browser to be more proactive in warnings.\\
	The Web browser is considered as the battlefield when fighting Malware, as most of the Internet traffic is channeled via the browser. There are many types of Malware that are interested in our surfing behavior and what we write online. These browser hijackers are usually qualified as spyware or Trojans. Other Malware - called hijackers - may take the user to sites of their choice. To increase the probability of a successful attack, the malicious code often checks the environment of the victim's (target) system, such as the browser type, browser plugins, Cookies, JavaScript features or version information.
	
	Such reconnaissance enables the attacker to find all available browser vulnerabilities to exploit. As observed in Wang et al. \cite{Wang_HoneyMonkeys_2006}, attackers often target multiple vulnerabilities during one attack in order to increase their chance of success. All this is done over a period of time without the user's knowledge.
	
	Of late, browser functionality has been enhanced to fight attacks, but more needs to be done in this field. Browser designers have kept user safety as their utmost goal while improving the user-Interface features. Using a safer browser may not be the best and final solution, since the more popular your browser is, the bigger the chance that some Malware author is looking for a security breach. Browsers need to extend safe browsing awareness into local languages. Another big contribution to how safe we are is how we behave online. The easier for the user to be tempted into clicking on anything online, the bigger the chance that at some point, the user will be a victim.\\
		
	2. Preload anti-Malware with the Operating System \\
	No mobile or tablet Operating System is completely safe, and protection is a requirement across such platforms. Since the Android platform is written on Java, Malware code can be written on a standard PC and transferred to the mobile or tablet platforms. Since many Android security Apps combine anti-theft features with backup and antivirus, it doesn’t hurt for the device vendors to pre-install them rather than leave it up to the user to shop around. Such Apps proactively scan all other downloaded Apps, as well as conduct periodic scans for potential threats. These Apps notify users if it identifies a new App as "potentially harmful" or suspicious, and will provide instructions on how to deal with them. Recently Kaspersky Lab has teamed up with Qualcomm to pre-install its antivirus app software on Snapdragon-powered Android devices.\\
	
	3. Increase training and awareness.\\
	Humans are the weakest link in security, and thus awareness trainings must be treated as continuous effort and just not a one-time activity. Online scammers often use opportunities such as tax returns seasons, holidays, social engineering, helpdesk support etc. to trick lay people into giving out their private information or allowing them to remote access their computers. Trust, Curiosity, Ignorance needs to be the focal point of all web-security awareness trainings. Internet users need to recognize the signs of Malware attack, like, a barrage of pop-up windows, a hijacked browser, sudden or repeated change of Internet browser's home page, appreance of unwanted or unexpected browser toolbars, random error messages, identifying ransom notes, and sluggish performance on the user's computer. Website reputation scores is gaining ground these days and can also be found as ad-ons to few web browsers \cite{Google_safe_browsing}, \cite{Wot_Chrome}, \cite{Wot_Mozilla}. \\
	
	4. Criminal prosecution of the Malware authors.\\
	There has been little cooperation among countries to prosecute Malware authors operating out of their soil. Also, with governments trying to position Malware as a cyber-attack weapon, it is difficult to prosecute Malware authors as they seem to have the protection of their country’s political systems. Although Malware criminals are difficult to apprehend; once caught, nearly all of them admit guilt \cite{Pompon}.
	
	In most Malware cases, an FBI investigation leads out of the United States. During these investigations, the FBI relies on their Legal Attachés or ‘legats’, who coordinate with other countries’ law enforcement officials. In countries where the police cooperate with the US authorities and are competent, the FBI is very open to allowing a local prosecution of a criminal. The danger is that without sufficient proven damages, judges in the U.S. can perceive the defendant as a ‘troubled smart kid’ and let them off with a light sentence. Sentences with high restitution also serve to punish the criminal and help defray the damages that victims suffer. In Malware cases, prison terms average 17 months, and restitution repayments average \$100,000 [6].\\
	
	5. Privacy watch and awareness\\
	These days, Internet or online privacy can be challenging to achieve given the easy access to Internet and constant use of Smartphones. Why privacy matters can be a long discussion, but anyone who values their online privacy should be proactive in safeguarding it. People need to limit sharing their personal information online or on social media as such information can be easily misused by nefarious actors. Companies track user's online behavior across websites to serve relevant advertisements. Internet users should always be watchful about Tracking, Surveillance and Theft of their identity and private data. There are many smart ways to help protect your privacy online such as; 
		\begin{enumerate}
		\item Browse in incognito or private mode. Delete Cookies when existing a browser and visit the privacy settings on browsers.
		\item Use a search engine that champion privacy.
		\item Update and apply patches against your applications, devices and computer whenever updates and patches are released by vendors. If possible, allow for automatic updates and patches. This is different than device "upgrade" which may involve cost and can be undertaken at a slower pace.
		\item Use a virtual private network.
		\item Watch where you click on websites.
		\item Secure your mobile devices using strong disk-level encryption if already not done so.
		\item Use quality antivirus software. Configure antivirus for automatic updates and scan periodically.
		\item Use Firewalls (application or network).
		\item Secure home network routers with strong encryption, set strong passwords and frequently recycle passwords.
		\item Adjust your social media privacy settings (Facebook, Twitter etc.).
		\item Secure your online/cloud storage and use encrypted connections to websites.
		\item Securely dispose electronic equipment.
	\end{enumerate}

	Securing Privacy can be subjective and each Internet user may term it differently. While users do not want to completely lock themselves out of the Internet world, they need to strike a balance between the careful use of the Internet and safeguarding their privacy.\\
		
	6. Securing Code Mashup, Maintenance, and Securing Systems\\
	Given the prevalence of client-side code mashups, it is imperative to design a sound approach to enhance the flexibility of the current mashup programming practice with guaranteed security \cite{Chang_2013}. Since not all the applications on the Internet are well designed and built, coding standards vary by programmers, causing exploits from the Malware authors who specialize in the drawbacks of the programming language or logic. A McAfee report found that more than 90\% of Malware attacks with the username/password combination of “admin:admin” originated from the United States \cite{McAfee}. Such lax security configurations of routers, IoT devices, network systems, etc. should be avoided. A proposed method is for the vendors of such systems to default their system configurations to a strong setting. Alternatively, application/functionality security settings logic should be strictly enforced by vendors to avoid weak configurations. Regular patching and upgrades is always recommended.
	
	\section{Conclusion}
	Curiosity, Ignorance, and Ignorance are the cornerstones of the Malware industry. They are the opium for Malware authors. Unfortunately, computer literacy does not bring about smart Internet surfing. With the growing population of new Internet users in the developing world who are not yet very computer savvy, it is easy for Malware authors to use such machines for Botnet attacks. Based on the browsing habits of a user, it would be beneficial to predict and warn if they are prone to Malware attack (high risk). Such commercial and affordable security controls, possibly driven by Artificial Intelligence, Deep Learning and Machine Learning, can be valuable in attack prediction. In the smartphone world, with Android security yet to be fully fortified, hundreds of Malware thrive as the common user does not care to install anti-Malware Apps and blindly trust App stores to keep out the malicious Apps. Users, especially in the developing world, where the Internet is fast catching up, do not seem to understand and value their privacy rights. In the developing world, victims seldom report Malware crimes as they do not value their personal data or do not expect any action from the online community or their governmental authorities. This only severely limits the apprehension of Malware criminals but degrades the overall effort of law enforcement to predict and punish Malware authors. User privacy, as seen in the developed world is still not a priority in the developing world. Thus Malware authors capitalize on this gap by catering differently to the developing and developed world. Malware industry adapts and shifts to its needs as in any other criminal industry. With Malware authors being now directed by governments to fight Cybersecurity, it's bound to grow unless netizens realize and educate themselves to safe browsing tactics. While human curiosity may be difficult to curb online, ignorance and trust can be easily treated with periodic awareness and training. \\

	\textit{Disclaimer: This publication was prepared by the author (Sundar Krishnan) in his personal capacity. The views and opinions expressed in this publication are solely those of the author. They do not purport to reflect the opinions, policies or views of the Department of Computer Science at Sam Houston State University or that of Sam Houston State University.}
	
	\bibliographystyle{IEEEtran}
	\bibliography{Malware}

	% that's all folks
\end{document}